\RequirePackage{fix-cm}
\documentclass[smallextended]{svjour3}       
\smartqed  
\usepackage{graphicx}
\usepackage{amsmath}
 \usepackage{mathptmx}      
\usepackage{latexsym}
\usepackage{physics}
\newcommand{\Beq}{\begin{equation}}
\newcommand{\Eeq}{\end{equation}}
\renewcommand{\vec}[1]{\boldmath{#1}}

\begin{document}

\title{Solution of Relativistic Feshbach-Villars Spin-1/2 Equations
}


\author{  D.~Wingard
\and   A.~Garcia~Vallejo
\and Z.~Papp$^{*}$
}


\institute{  D.~Wingard
	\at
              Department of Physics  and Astronomy \\
              California State University Long Beach \\
              Long Beach, California, USA \\
              \email{dwingard@uw.edu}     
\and
            A.~Garcia~Vallejo
	\at
		Department of Physics  and Astronomy \\
		California State University Long Beach \\
                Long Beach, California, USA \\
          and\\
               Department of Physics \\
               New Mexico State University\\
               Las Cruces, New Mexico, USA \\
              \email{antoniov@nmsu.edu}
 \and             
                Z.~Papp$^{*}$
	\at
              Department of Physics  and Astronomy \\
              California State University Long Beach \\
              Long Beach, California, USA\\
              \email{Zoltan.Papp@csulb.edu} 	      
}

\date{Received: date / Accepted: date}

\maketitle

\begin{abstract}
We propose method for studying relativistic spin-$1/2$ particles by solving the corresponding 
Feshbach-Villars equation. We have found that the Feshbach-Villars spin-$1/2$ equations can be formulated 
as spin-coupled Feshbach-Villars spin-$0$ equations,
that results in a Hamiltonian eigenvalue problem.
We adopted an integral equation formalism.
The potential operators are represented in a discrete Hilbert space basis
and the relevant Green's operator has been calculated by a matrix continued fraction.

\end{abstract}

\section{Introduction }

The basic equations of relativistic quantum mechanics, the Klein-Gordon and the Dirac equations are quite different in form
and they also differ considerably from the non-relativistic Schr{\"o}dinger equation. 
The Schr\"odinger equation is linear in time derivative and quadratic in spatial derivatives. On the other hand the Klein-Gordon
equation is quadratic, while the Dirac equation is linear in both type of derivatives. 

In 1958, by ``squaring''  the Dirac equation, Feynman and Gell-Mann  proposed a variant 
that is quadratic in both types of variables \cite{feynman1958theory}.  About the same time, Feshbach and Villars linearized 
the Klein-Gordon in
the time variable \cite{Feshbach:1958wv}. As a result, the Feshbach-Villars equations are linear in time and
quadratic in spatial derivatives, just like the Schr\"odinger equation, but the wave function has two components.  
Much later in 1996, Robson and Staudte introduced the Feshbach-Villars linearized form of the Feynman--Gell-Mann equation
\cite{robson1996eight,staudte1996eight}.
The Feshbach-Villars linearization of the relativistic equation results in a Hamiltonian that is not Hermitian in the usual sense,
yet it possesses  real eigenvalues. The Hamiltonian is multicomponent and explicitly exhibits the particle-antiparticle features.

Although the Feshbach-Villars formalism has some history, it has been used rather scarcely in actual calculations. 
The reason maybe is that the multicomponent non-Hermitian feature makes the solution rather cumbersome.
Nevertheless, there are some notable recent exceptions 
\cite{guettou2006pair,robson2001relativistic,merad2000boundary,khounfais2004scattering,boudjedaa1999exact,boudjedaa2000path,merad2001path,benzair2013noncommutative,boudjedaa2019fermionic,boudjedaa2020bosonic,BilelMerad2020,BOUZENADA2023116288,BOUZENADA2023169479,BOUZENADA2023169302,Silenko:2020gpv,Silenko_2020}.

The aim of this paper is to solve the Feshbach-Villars equation (FV1/2) for spin-1/2 particles. This work is the continuation of
our previous one where we solved the spin-0 (FV0) equation \cite{brown2016matrix,motamedi2019relativistic}. 
In Sections \ref{sec2} and \ref{sec25} we 
show the Feynman--Gell-Mann and Feshbach-Villars equations for spin-1/2 particles, respectively. 
In Section \ref{sec3} we present the solution method
as the extension of method in Ref.\  \cite{motamedi2019relativistic}. The numerical power of the method 
is illustrated in Section \ref{sec5} and conclusions are given  in Section \ref{sec6}.

\section{Feynman--Gell-Mann equation  }
\label{sec2}

Consider a relativistic particle with charge $q$ in electromagnetic field $A_{\mu} =(\vec{A},i\Phi)$. The Dirac 
equation for the four-component wave function $\psi$  reads
\Beq
\left[ \gamma_{i} \left( \pdv{x_{i}} - \frac{iq}{\hbar c} A_{i}\right) +  \frac{m c}{\hbar}  \right] \psi=0,
\Eeq
where $\gamma_{i}$ are the $4\times 4$ gamma matrices satisfying the anti-commutation relation
\Beq
\comm{\gamma_{i}}{\gamma_{k}}_{+} = \gamma_{i} \gamma_{k}+ \gamma_{k} \gamma_{i}= 2 \delta_{ik}.
\Eeq

Feynman and Gell-Mann 
``squared''  this equation
\Beq
\left[ \gamma_{k} \left( \pdv{x_{k}} - \frac{iq}{\hbar c} A_{k}\right) -   \frac{m c}{\hbar} \right] 
 \cdot \left[ \gamma_{i} \left( \pdv{x_{i}} - \frac{iq}{\hbar c} A_{i}\right) + \frac{m c}{\hbar}   \right]   \psi=0. 
 \label{feynman58}
\Eeq
With a possible representation of gamma matrices,
\Beq
\gamma_{1}  =\left(\begin{array}{@{}rrrc@{}}
0&0&0&-i\\
0&0&-i&0\\
0&i&0&0\\
i&0&0&0
\end{array}\right), \ 
\gamma_{2}=\left(\begin{array}{@{}rrrc@{}}
0&0&0&-1\\
0&0&1&0\\
0&1&0&0\\
-1&0&0&0
\end{array}\right), \ 
\gamma_{3} =\left(\begin{array}{@{}rrrc@{}}
0&0&-i&0\\
0&0&0&i\\
i&0&0&0\\
0&-i&0&0
\end{array}\right), \ 
\gamma_{4}=\left(\begin{array}{@{}rrrc@{}}
1&0&0&0\\
0&1&0&0\\
0&0&-1&0\\
0&0&0&-1
\end{array}\right),
\Eeq
Eq.\ (\ref{feynman58}) becomes
\Beq
\begin{split}
\left(D_{1}^{2} + D_{2}^{2}+D_{3}^{2}+D_{4}^{2} -   {m^{2} c^{2}}/{\hbar^{2}}  \right) \psi +
 \frac{q}{\hbar c} ({\cal B} \vec{\sigma} - i {\cal E} \vec{\sigma}) \psi & = 0 \\
\left(D_{1}^{2} + D_{2}^{2}+D_{3}^{2}+D_{4}^{2} - {m^{2} c^{2}}/{\hbar^{2}}  \right) \psi + 
\frac{q}{\hbar c} ({\cal B} \vec{\sigma} - i {\cal E} \vec{\sigma}) \psi & = 0, 
\end{split}
\label{2feynmanGell}
\Eeq
where $\vec{\sigma}$'s are the $2\times 2$ Pauli spin matrices,
\Beq
\sigma_{1}= \mqty(0 & 1 \\ 1 & 0), \quad \sigma_{2}= \mqty(0 & -i \\ i & 0), \quad \sigma_{3}= \mqty(1 & 0 \\ 0 & -1), 
\Eeq
${\cal B}=\nabla \cross \vec{A}$ is the magnetic field and 
${\cal E}=-1/c \pdv*{\vec{A}}{t} - \nabla \Phi$ is the electric field.
We thus have two identical two-component equations and
 can drop one of them to work with a two-component formalism.
Instead of the four-component first order Dirac equation we can have, without loss of generality,
 the two-component second order Feynman--Gell-Mann equation. By ``squaring'' the Dirac equation, 
 Feynman and Gell-Mann doubled
 the solution space, which is reduced again by keeping only one of Eq.~\eqref{2feynmanGell}.
 
 \section{Feshbach-Villars equation for spin-1/2 particles }
\label{sec25}

In this work we consider a stationary field with  $\vec{B}=0$ and assume that $q\Phi=V$. 
Then one of Eqs.~(\ref{2feynmanGell}) becomes
\Beq
\left(-\hbar^{2} c^{2} \nabla^{2} - (i \hbar \pdv*{t}-V)^{2} - m^{2}c^{4} + i \hbar c \nabla V \vec{\sigma} \right) \psi =0,
\Eeq
or
\Beq
(i \hbar \pdv*{t}-V)^{2} \psi = \vec{p}^{2}c^{2} \psi + m^{2}c^{4} \psi - i \hbar c  \nabla V \vec{\sigma} \psi.
\Eeq
The Feshbach-Villars linearization amounts of splitting the wave function into components such that
\Beq
\psi = \phi  + \chi \quad \text{and} \quad (i \hbar \pdv*{t}-V) \psi = mc^{2}(\phi-\chi). 
\Eeq
This leads to the set of equations
\Beq
\begin{split}
 (i \hbar \pdv*{t}-V) (\phi + \chi) & = mc^{2} ( \phi -  \chi) \\
  (i \hbar \pdv*{t}-V) (\phi - \chi) & = \frac{\vec{p}^{2}}{m} (\phi + \chi) + mc^{2}  (\phi + \chi) 
   - \frac{i\hbar}{mc} \nabla V  \vec{\sigma} (\phi + \chi).
\end{split}
\Eeq
Reorganizing, we obtain a Schr\"odinger-like Feshbach-Villars equation for spin-1/2 particles
\Beq
i \hbar \pdv{t} \mqty( \phi \\ \chi ) = H \mqty( \phi \\ \chi ),
\Eeq
with Hamiltonian
\Beq 
{H} = \mqty(1 & 1 \\ -1 & -1) \frac{\vec{p}^{2}}{2m}  + \mqty(1 & 0 \\ 0 & -1) mc^{2} 
+ \mqty(1 & 0 \\ 0 & 1)V -    \mqty(1 & 1 \\ -1 & -1) \frac{i\hbar}{2mc} \nabla V  \vec{\sigma}.
\Eeq
With the help of Pauli and the unit matrices, that act in the Feshbach-Villars component space,
\Beq
\tau_{1}= \mqty(0 & 1 \\ 1 & 0), \quad \tau_{2}= \mqty(0 & -i \\ i & 0), \quad \tau_{3}= 
\mqty(1 & 0 \\ 0 & -1),  \quad I_{2}= \mqty(1 & 0 \\ 0 & 1),
\Eeq
we can write the Hamiltonian as
\Beq 
{H} =  (\tau_{3}+i \tau_{2}) \frac{\vec{p}^{2}}{2m}  + \tau_{3} mc^{2} 
+  I_{2}V -  (\tau_{3}+i \tau_{2})  \frac{i\hbar}{2mc} \nabla V  \vec{\sigma}.
\Eeq

We could also introduce a scalar interaction into the formalism by the substitution $m \to m + S/c^{2}$. 
This results in the Hamiltonian
\Beq 
{H} =  (\tau_{3}+i \tau_{2})   \left( \frac{\vec{p}^{2}}{2m} +U \right) + \tau_{3} mc^{2} 
+ I_{2} V -    (\tau_{3}+i \tau_{2})  \frac{i\hbar}{2mc} \nabla V  \vec{\sigma},
\label{HUV}
\Eeq
where $U= S + S^{2}/2mc^{2}$.

\section{The solution method }
\label{sec3}

Here we assume that both $V$ and $U$ are spherical potentials and thus depend only on radial variable $r$.
Then the Hamiltonian forms a complete set of commuting 
observables with angular momentum operator $J^{2}$ and $J_{z}$. 
The common eigenstates of $J^{2}$ and $J_{z}$ are the spin-orbit coupled angular momentum states 
\Beq
\Phi^{(\pm)}_{j,m}(\vartheta,\varphi)  =
 \sum_{m_{l}, m_{s}} \langle l_{\pm} , 1/2 ; m_{l}, m_{s} | j, m \rangle Y_{l_{\pm}} (\vartheta,\varphi)  \chi_{1/2, m_{s}},
\Eeq
where $j = l_{+}+1/2 = l_{-}-1/2$ and $ \chi_{1/2, m_{s}}$ denotes the spin states. 
For a spherical potential $V$ the spin coupling term  reads
\Beq
H' = - (\tau_{3}+i \tau_{2})  \frac{i\hbar}{2mc}  \dv{V(r)}{r} \vec{e}_{r}  \vec{\sigma},
\Eeq
where $\vec{e}_{r}$ is a unit vector in the radial direction.
It has been shown in Refs.\ \cite{auvil1978relativistic,holstein2013topics,wingard2023} that
\Beq
\vec{e}_{r}  \vec{\sigma} \ket{ \Phi^{(\pm)}_{j,m} } = \ket{ \Phi^{(\mp)}_{j,m} },
\Eeq
i.e.\  this term couples the basis states with different orbital angular momenta.

As a consequence,  the angular momentum projected Hamiltonian becomes
\Beq
H = \mqty( \tilde{H}^{(+)}_{FV0}  & 0 \\ 0 & H^{(-)}_{FV0} ) + 
\mqty( 0  & \tilde{H}'^{(+-)} \\ \tilde{H}'^{(-+)} & 0 ) ,
\Eeq
where
\Beq
 \tilde{H}^{(\pm)}_{FV0} = \mel{\Phi^{(\pm)}_{j,m}}{ (\tau_{3}+i \tau_{2})   
 \left( \frac{\vec{p}^{2}}{2m} +U \right) + \tau_{3} mc^{2} 
+ I_{2} V }{\Phi^{(\pm)}_{j,m}} 
\label{hfv0}
\Eeq
and
 \Beq  
  \tilde{H}'^{(\pm \mp)} =    - (\tau_{3}+i \tau_{2})  \frac{i\hbar}{2mc} \mel{\Phi^{(\pm)}_{j,m}}{\dv{V(r)}{r} }{\Phi^{(\mp)}_{j,m}}.
  \label{hpfv12}
\Eeq
It should be noted that $\tilde{H}^{(\pm)}_{FV0}$ and   $\tilde{H}'^{(\pm \mp)}$ are still operators in radial variables.

We can solve this problem in the way presented in Ref.\ \cite{motamedi2019relativistic}. Assume that potentials 
$U$ and $V$ are combinations of long range and short range terms and write the eigenvalue problem in a 
Lippmann-Schwinger form.
We have to put the long range potentials that behave asymptotically like $1/r$ for Coulomb, 
or like $r$ or $r^{2}$ for confining systems, into the Green's operator.
We can represent the short range potentials in a Hilbert space basis, in particular, in a Coulomb-Sturmian basis. 
The matrix elements of the short range potentials in 
Eqs.\ (\ref{hfv0})  and  (\ref{hpfv12}) can always be evaluated, at least numerically. 
The corresponding matrix elements of the long range  Green's operator can be calculated as a matrix continued fraction.
Then,  the whole problem becomes a linear algebraic problem with a zero search of a determinant.

\section{Numerical Illustrations}
\label{sec5}

To illustrate the power of the method we take the same example we had in Ref.\ \cite{motamedi2019relativistic}.
We adopt units such that $m=\hbar=e^{2}=1$ and $c=137.036$ and the potential is given by
\Beq
V(r) = 92/r - 240 \exp(-r)/r + 320 \exp(-4r)/r.
\label{couls}
\Eeq
This potential is Coulomb plus short range type, used typically in atomic and nuclear physics calculations.

\begin{table}[h!]
\centering
\caption{Bound and resonant state energies in potential of Eq.\ (\ref{couls}) for $l=0$.} 
\label{table1}
\begin{tabular}{ c c c  }
\hline
    Non-relativistic &  FV0  &  FV1/2   \\
    \hline
 \,\,\,\,-5.929368&-5.933465&-5.928157 \\
 \,\, 15.60918&   15.59950   & 15.60266   \\
 \,\,\,\,\,\,\,\,\,\,\,\,\,\,\,\,\,\,-0.0000015 $\textbf{\textit{i}}$	&  \,\,\,\,\,\,\,\,\,\,\,\,\,\,\,\,\,\,\,\,\,\,\,\,\,  -0.00000000002 $\textbf{\textit{i}}$	 &   \,\,\,\,\,\,\,\,\,\,\,\,\,\,\,\,\,\,\,\,\,\,\,\,\,\,\,\, -0.000000000006 $\textbf{\textit{i}}$\\
  \hline
\end{tabular} 
\end{table} 

In another example we consider a scalar confinement potential and a Coulomb vector potential,
\Beq
U(r) = \alpha_{1} r + \alpha_{2}r^{2} \quad \text{and} \quad V(r) = Z/r.
\Eeq
Table~\ref{table2}  and \ref{table3}  show the non-relativistic and the relativistic FV0 and FV1/2 energies with $l=0$ and $l=1$.
We take Coulomb $V(r) = -1/r$ plus linear $U(r)=r$
and quadratic $U(r)=r^{2}/2$ confinement potentials, respectively. These are again typical short range plus confining potentials, they
are used to model quark-quark interactions. 

We can see from the numerical illustrations that the presented numerical solution of the FV equations are numerically sound and can
pinpoint even tiny relativistic effects.

\begin{table}
\centering
\caption{Bound states in Coulomb plus linear confinement potential ($Z=-1$ and $\alpha_{1}=1$). }
\label{table2}

\begin{tabular}{|l| l | l  | l  | l | l  |   }
\hline
   Schr (l=0)& FV0 (l=0) &FV1/2&Schr (l=1)  &  FV0 (l=1) &  FV1/2  \\
\hline
 0.577924 &  0.577749  &0.577806&1.974214 &  1.974013 & 1.974004 \\
 2.450164 &  2.449834 &2.449862&3.335497 & 3.335086 & 3.335080 \\
3.756907 &  3.756356 &3.756377&4.468114 & 4.467458 & 4.467453 \\
4.855672 &  4.854865 &4.854883& 5.472592 & 5.471664 & 5.471660 \\
 5.836031 &  5.834942 &5.834958&6.391709 & 6.390484 &  6.390481 \\
  6.736622 &  6.735228 &6.735242& 7.248384 &7.246845  &7.246841   \\

    \hline
\end{tabular} 

\end{table} 

\begin{table}
\centering

\caption{Bound states in Coulomb plus quadratic confinement potential ($Z=-1$ and $\alpha_{2}=1/2$). }
\label{table3}

\begin{tabular}{|l|l|l| l  | l | l  |   }
\hline
    Schr (l=0)&FV0 (l=0)&  FV1/2&  Schr (l=1)&FV0 (l=1) &  FV1/2   \\
\hline
 0.179668 &  0.179538 &0.179595&1.709018 & 1.708842 & 1.708831 \\
 2.500002 &2.499612 &2.499652& 3.801930 & 3.801378 &3.801368 \\
4.631955 &4.631087 &4.631123& 5.860357 &5.859219 & 5.859209 \\
 6.712598 &6.711039  &6.711074& 7.902318 & 7.900380 &7.900371 \\
 8.769522 & 8.767059 &8.767092&9.934707 &9.931758 &  9.931749 \\
   10.81293 & 10.80935 &10.80938& 11.96088 &11.95671 & 11.95670  \\

    \hline
\end{tabular} 

\end{table}

\section{Summary}
\label{sec6}

The Feshbach-Villars formalism offers a unification of non-relativistic and relativistic equations. The FV0 equation formally 
looks like a Schr\"odinger equation, with various $2\times 2$ matrices attached to the kinetic energy, the mass and the potential and
the wave function components are coupled by the kinetic energy operator. This equation explicitly manifests the underlying 
particle-aniparticle feature of the nature. The FV1/2 equation adds one more layer, the spin structure. Then FV1/2 is basically
two FV0 equations coupled by the derivative of the vector potential. 

In this work we have presented a numerical method that solves the Feshbach-Villars equations for spin-$1/2$ particles.
The method is a generalization of our previous work that solved the Feshbach-Villars equation for spin-$0$ particles. 
These equations are genuine relativistic equations, that explicitly reveal the particle-antiparticle features. 
We addressed the problem through evaluating the corresponding resolvent operator via matrix continued fraction. 
By adopting integral equation formalism we eliminated the need for boundary conditions for the multicomponent wave function.
The numerical procedure is solid and stable that it can give account the sometimes tine relativistic effects.


\bibliographystyle{spphys}       
\bibliography{fv00}

%
%

\end{document}